\documentclass[12pt,epsfig]{article}
\usepackage{amsmath}
\usepackage{graphicx}
\usepackage{epstopdf}
\usepackage{color}

\begin{document}

	\vskip.1in	
	\begin{center}
		\textbf {Unsupervised classification of eclipsing binary light curves through $k$-medoids clustering}\\
		\vskip.2in Soumita Modak$^{1,*}$, Tanuka Chattopadhyay$^{2}$\\
		{\small and}\\
		Asis Kumar Chattopadhyay$^{1}$\\
		\vskip.1in
		$^{1}$Department of Statistics, University of Calcutta\\35 Ballygunge Circular
		Road, Kolkata- 700019, India\\
		email: soumitamodak2013@gmail.com\\
		email: akcstat@caluniv.ac.in\\
		\vskip.1in
		$^{2}$Department of Applied Mathematics, University of Calcutta\\
		92 A.P.C. Road, Kolkata- 700009, India\\
		email: tanuka@iucaa.ernet.in\\
		\vskip.1in
	\end{center}
	\begin{abstract}
		This paper proposes $k$-medoids clustering method to reveal the distinct groups of 1,318 variable stars in the Galaxy based on their light curves, where each light curve represents the graph of brightness of the star against time. To overcome the deficiencies of subjective traditional classification, we separate the stars more scientifically according to their geometrical configuration and show that our approach outperforms the existing classification schemes in astronomy. It results in two optimum groups of eclipsing binaries corresponding to bright, massive systems and fainter, less massive systems.
	\end{abstract}
	\vspace{0.1in}\textbf{Keywords:} Light curve of variable star; Clustering; $k$-medoids method; Complexity invariance distance.\\
	\section{Introduction}
	Eclipsing binaries (Es) are treated as fundamental probe for studying stellar structure and evolution (Akerlof et al. 2000; Bradstreet and Steelman 2002; Street et al. 2004; Graczyk et al. 2011; Chattopadhyay et al. 2016). They can be classified based on their light curves (LCs) to find out the possible sources of homogeneous groups of the stars (Kochoska et al. 2017; Mowlavi et al. 2017; S\"{u}veges et al. 2017 and references therein). 
	Miller et al. (2010) have carried out observations of 1,318 new variable stars covering 0.25 square degree region of the Galactic plane centered on Galactic coordinates (latitude, longitude) of
	$(330.94,-2.28)$ deg. The majority of stars in the above region are thought to be associated with the Normal Spiral Arm. They separated the stars subjectively according to the appearance of their LCs into four groups, viz. Algol type (EA), Beta Lyrae (EB), W Ursae Majoris (EW) and un-categorized pulsating stars (PUL). But such subjectivity usually includes degeneracy, i.e. it classifies the stars with different physical properties in the same group. Hence, this traditional scheme is now almost obsolete and can be rather misleading. Most importantly, many of the stars were categorized with uncertainty or ambiguity. These limitations lead us to clustering  the stars objectively without assuming any prior information.
	
	In this paper, we carry out an unsupervised classification of the variable stars from Miller et al. (2010) by applying the $k$-medoids clustering method (Kaufman and Rousseeuw 2005) to their LCs. Astronomical data collection is often obscured by bad weather conditions. Usually, it is not possible to have repeated observations on astronomical objects due to intermediate celestial objects and instrumental restrictions. As a result we have data contaminated with noise, affected by outliers or sparsely distributed (Feigelson and Babu 2013; Modak et al. 2017, 2018; Bandyopadhyay and Modak 2018; Modak and Bandyopadhyay 2019), which cannot be analyzed  properly by the usual methods. To overcome such problems we apply the nonparametric partitioning based $k$-medoids clustering to univariate time series, i.e. LCs, which is robust against noisy or unusual LCs while facilitating any distance measure to extract relevant temporal information properly from the given LCs. Previous works of classification of Es based on LCs (Sarro et al. 2006; Malkov et al. 2007; Pr\v{s}a et al. 2008; Matijevi\v{c} et al. 2012; Kirk et al. 2016) suffer from many significant drawbacks, e.g. the methods are imposed with a lot of restrictions or model assumptions, and the classification is supervised in the sense that the number of classes is assumed as a priori or the properties of the classes are known. Some performed the classification 
	based on variables derived from the observed or simulated LCs instead of directly using the observed LCs of stars. Their resulting classes overlap significantly with respect to the properties of the stars. However, our method can be used for any new data set where the distribution of the LCs is unknown and assumptions on the number of clusters or the cluster properties are not possible to make before the analysis. The present paper uses the $k$-medoids method with complexity invariance distance (CID) (Prati and Batista 2012; Batista et al. 2014; Wei 2014), dynamic time warping distance (DTW) (Rabiner and Juang 1993; Keogh and Ratanamahatana 2005; Giorgino 2009; Cassisi et al. 2012) and Euclidean distance (ED), where CID gives the best clustering in terms of the average silhouette width (ASW) (Rousseeuw et al. 1987) with two resulting groups of Es. We prove the superiority of our approach over DBSCAN method (Ester et al. 1996; Kochoska et al. 2017), $t$-SNE technique (Maaten 2014; Kirk et al. 2016; Kochoska et al. 2017) and $k$-means method (Hartigan and Wong 1979; Modak et al. 2018). The success of the proposed method in revealing distinct groups of stars is also confirmed by fuzzy $k$-means clustering (Bezdek 1981). Simulation study establishes the usefulness of our approach in general as a potential LC-based classifier.
	
	The paper is organized as follows. In Section 2, we describe the data and the transformations on data. Section 3 discusses the clustering method with accuracy measure, distance measures and the simulation study. Result and discussion are presented in Section 4, whereas Section 5 concludes.
	\section{Data}
	The present data set is taken from Miller et al. (2010) where each of the 1318 variable stars in our Galaxy has a LC file together with R-band magnitude (R), colors (B-R, R-I) and period (P). From these we obtain the variables I as R minus R-I, B-I as B-R plus R-I and B as B-I plus I. According to the subjective classification, the variable stars mainly consist of Es along with some possible pulsating stars (201 uncertain and 118 potential pulsating stars). For each LC, relative flux variation in R-band is given on a continuous time scale in Heliocentric Julian Date (HJD) within the range from 2452450.622 HJD to 2453607.616 HJD. Each LC is unevenly spaced of different length having values at different time points. The length of LCs varies from 130 to 264 (except one LC having length 5) and period ranges from several hours to several weeks.
	\subsection{Phase computation, interpolation and binning}\label{phase}
	Comparison of the  given LCs is performed in terms of a full cycle over phase interval [0,1], which restores the physical properties of the stars (Percy 2007; Deb and Singh 2009; Soszy\'{n}ski et al. 2016), having observations at $l$ (say) evenly spaced phase points, obtained as follows.\\
	(i) We obtain phased LCs by transforming the given time points into standard phases (always lie between 0 and 1) using the following equation (Percy 2007; Deb and Singh 2009)
	\begin{equation}\label{standardphase}
	\text{decimal}\hspace{1 mm}\text{portion}\hspace{1 mm}\text{of}\hspace{1 mm}[(t-t_{0})/\text{P}],
	\end{equation}
	where $t$ denotes the time of measurement of the star (here in HJD),
	$t_{0}$ is an arbitrary epoch - usually a time of maximum or minimum brightness (here the time of the first observed maximum) and
	P represents the period (known and constant) of the star (here in days). 
	Now, we have the $i^{th}$ LC over the phase interval $[0,p_{i}]$, where $p_{i}$ (close to 1 but $<$1) is maximum of standard phases for the $i^{th}$ LC, $i = 1,\ldots,1318$ (here $p_i$'s are different for all $i$).\\
	(ii) To avoid extrapolation involving larger uncertainty than interpolation, we extend the phase interval of the $i^{th}$ LC to $[0,p_i+1]\in[0,2)$, $i=1,\ldots,1318,$ by adding $+1$ to the standard phases computed in step (i) (since a phase of 0 is the same as a phase of 1, -1 or 2).\\
	(iii) We fit the linear spline (Press et al. 1992; Cassisi et al. 2012) to the LCs from step (ii) at $l$ evenly spaced phases over [0,1]. Given a tabulated function $y_{i}=y(x_{i})$, $i =1,\ldots,l_i$ with $x_{i}<x_{i+1}$ for $ i = 1,\ldots,l_i-1$, the interpolating function joins $l_i-1$ linear functions of the form $f_{i}(x)=a_{i}y{_i}+b_{i}y_{i+1}$, $x\in[x_{i}, x_{i+1}$], where $a_{i}$, $b_{i}$ are constants satisfying (a) $f_{i}(x_{i})=y_{i}$, (b) $f_{i}(x_{i+1})=y_{i+1}$, i.e. $a_{i}=\frac{x_{i+1}-x}{x_{i+1}-x_{i}}$ and $b_{i}=1-a_{i}=\frac{x-x_{i}}{x_{i+1}-x_{i}}$, $i=1,\ldots,l_i-1$.\\
	(iv) The interpolation error in step (iii) decreases with the increase in $l$, provided $l$ should not be too large to cause considerable error for under-sampled series. Here the LCs (except one) are well sampled of length at least 130 with most of them of length close to $272$ which is our chosen value of $l$ such that there is no loss of information, significant computational burden or considerable error in approximating the series of lower length. Technically, $l$ is varied over different plausible values which robustly gives the optimal number of clusters as two in terms of clustering accuracy ASW (see, Section \ref{ON}) with $l=272$ corresponding to the best clustering.
	\section{Classification scheme}
	Clustering of time series (Liao 2005; Liao et al. 2006), either in time domain or frequency domain (Dargahi-Noubary 1992; Caiado et al. 2006), can be performed on series of equal length or unequal length (Lomb et al. 1976; Caiado et al. 2009; Stefan et al. 2013), evenly spaced or unevenly spaced (Scargle 1989; Moller-Levet et al. 2003; Eckner 2014, 2017). Here we apply the following unsupervised classification scheme to the evenly spaced LCs of equal length over the phase interval [0,1], which are referred to as `objects' in the cluster analysis.
	
	\subsection{$k$-medoids: Partitioning Around Medoids (PAM)}
	$k$-medoids is a nonparametric partitioning based clustering method which can be applied to univariate time series like LCs. We use a fast and efficient algorithm `PAM' (Kaufman and Rousseeuw 2005) which is executed using the inbuilt function `pam' in software `R'. It is based on the search for $k$ medoids in the data set, where a medoid is the representative object of the cluster it belongs to. These $k$ medoids represent the various structural aspects of the data set of size $N$ being partitioned into $k$ mutually exclusive and exhaustive clusters around $k$ medoids, where a medoid is that object of the cluster for which the sum of distances to all other objects of the cluster is minimal. Because of the medoids this method is robust against noise, outliers or sparsely distributed data (Singh and Chauhan 2011). This clustering method allows any distance measure depending upon the nature of the given data. 
	\subsection{Competitive methods of unsupervised clustering for comparison}
	As a competitor of $k$-medoids clustering we consider the very popular distance-based approach DBSCAN method (Ester et al. 1996; Kochoska et al. 2017) with parameters `$\epsilon$' and `$MinPts$' which estimates the density around each object by counting the number of objects in a neighborhood ($\epsilon$) and applies a threshold ($MinPts$) to identify core,
	border and noise objects. The core objects form a cluster if there is a chain of them wherein one falls inside the $\epsilon$-neighborhood of the next, whereas the border objects are arbitrarily assigned to the clusters. Objects lying outside the $\epsilon$-neighborhood of the clustered objects cannot be assigned to a cluster and is considered as noise. We also apply this DBSCAN method to data transformed into a nonlinear form through $t$-SNE (Maaten 2014; Kirk et al. 2016; Kochoska et al. 2017) which reduces high-dimensional objects in a low-dimensional space, where similar objects are modeled using close transformed data and dissimilar objects are modeled using far transformed data with high probability. Here any appropriate metric can be chosen to compute distance among the objects. 
	
	Another classical clustering $k$-means (Hartigan and  Wong 1979) is well-known to partition the objects into $k$ clusters in which each object belongs to the cluster with the nearest mean, whereas fuzzy $k$-means (Bezdek 1981) groups the data set such that each object can be classified into more than one cluster. Here for every object the degree of being member of a cluster is quantified by a membership $(\in[0,1])$, with larger membership value indicating higher probability that the object inherits the properties of that cluster. In fuzzy $k$-means, the centroid of a cluster is the weighted mean of all objects where weights are the power function of membership values belonging to the cluster.
	
	\subsection{Optimal number of clusters}\label{ON}
	The optimal value of the number of clusters ($k$) is chosen from ASW (Rousseeuw et al. 1987), which accounts for the efficacy of the cluster analysis, using the distance measure appropriate for the given objects. For each object $i$, the silhouette width (SW) $s(i)$ lies from -1 to 1. Object with a large positive SW is very well clustered, SW around 0 means that the corresponding object lies between two clusters, and object with a small negative SW is probably placed in the wrong cluster. ASW is the average SW over all $i$, with $-1\leq$ ASW $\leq1$, which is calculated for $k=2,3,\ldots,$ etc. and the value of $k$ is chosen for which ASW is maximum. For given $k$, the silhouette plot (Rousseeuw et al. 1987) gives a graphical representation of SW of each object. Robustness of the optimal number of clusters with respect to clustering accuracy measure is verified by another distance-based quantity, the connectivity (Handl et al. 2005), which indicates the degree of connectedness of the clusters and takes a value between zero and infinity with a minimum corresponding to the best possible clustering.
	\subsection{Distance measure}\label{DM}
	Time series like the LCs having a large number of peaks, in different quantities, amplitudes or durations are considered as complex time series. So we measure distances between the LCs under complexity invariance using CID (Prati and Batista 2012; Batista et al. 2014; Wei 2014). Because Batista et al. (2014) empirically showed that CID generally performs best among its possible competitors and is effective in clustering complex time series. Here the LCs show considerable complexity in terms of the complexity estimate defined in Eq. \eqref{CE}.
	CID between two time series $X$ with values $x_{1},\ldots,x_{n}$ and $Y$ with values $y_{1},\ldots,y_{n}$ corresponding to time points $t=1,\ldots,n$, is defined as
	\begin{equation}
	\text{CID}(X,Y)={\text{ED}}(X,Y)\times {\text{CF}}(X,Y),
	\end{equation}
	where CF is a complexity correction factor given by
	\begin{equation*}
	\text{CF}(X,Y)=\frac{\max({\text{CE}}(X),{\text{CE}}(Y))}{\min({\text{CE}}(X),{\text{CE}}(Y))},
	\end{equation*}
	with the following complexity estimate of time series $X$
	\begin{equation}\label{CE}
	\text{CE}(X)=\sqrt{\sum\limits_{t=1}^{n-1} (x_{t}-x_{t+1})^2};
	\end{equation}
	and
	\begin{equation*}
	\text{ED}(X,Y)=\sqrt{\sum\limits_{t=1}^n (x_{t}-y_{t})^2}.
	\end{equation*}
	
	We consider another popular time series distance DTW (Rabiner and Juang 1993; Keogh and Ratanamahatana 2005; Giorgino 2009; Cassisi et al. 2012).
	For two time series $X$ with values $x_{1},\ldots,x_{m}$ and $Y$ with values $y_{1},\ldots,y_{n}$ corresponding to time points $t=1,\ldots,m$ and $t=1,\ldots,n$ respectively, dynamic time warping finds the warping path $W={w_{1},\ldots,w_{l},\ldots,w_{L}}$ of contiguous elements on the local distance matrix having $(i,j)^{th}$ element  $d(x_{i},y_{j})=|x_{i}-y_{j}|$ $(i=1,\ldots,m,j=1,\ldots,n)$, such that $w_{l}=(i_l,j_{l})\in\{1,\ldots,m\}\times\{1,\ldots,n\},l=1,\ldots,L,\max(m,n) \leq L < m+n -1$, satisfy the following conditions.
	(C1) Boundary conditions: $w_{1} = (1,1)$, $w_{L} = (m, n)$,
	(C2) Continuity: For $w_{l+1}=(i_{l+1},j_{l+1})$ and $w_{l}=(i_{l},j_{l})$, $i_{l+1}-i_{l}\leq1$ and $j_{l+1}-j_{l}\leq1$ for all $l=1,\ldots,L-1$ and
	(C3) Monotonicity: For $w_{l+1}=(i_{l+1},j_{l+1})$ and $w_{l}=(i_{l},j_{l})$, $i_{l+1}-i_{l}\geq0$ and $j_{l+1}-j_{l}\geq0$ for all $l=1,\ldots,L-1$. Then,
	DTW which is an optimal
	path between $X$ and $Y$ under the stated restrictions is defined as
	\begin{equation*}
	{\text{DTW}}(X,Y)=\min\bigg(\sqrt{\sum\limits_{l=1}^L w_{l}}\bigg).
	\end{equation*}
	Dynamic programming is used to find this path by evaluating the following recursive function (Giorgino 2009),
	\begin{equation*}
	\begin{split}
	g[i,j]  = & \min\bigg(g[i,j-1] + d(x_{i},y_{j}),g[i-1,j-1] + 2\hspace{.05 in} d(x_{i},y_{j}),\\
	&g[i-1,j] + d(x_{i},y_{j})\bigg),\hspace{.07 in} i=1,\ldots,m,j=1,\ldots,n.
	\end{split}
	\end{equation*}
	In our study, $m=n$ wherein ED comes as a particular form of DTW with $w_l=(i_l,j_l),i=j=l$.
	
	\subsection{Simulation study}
\begin{figure}
		\centering
		\includegraphics[width=1\textwidth]{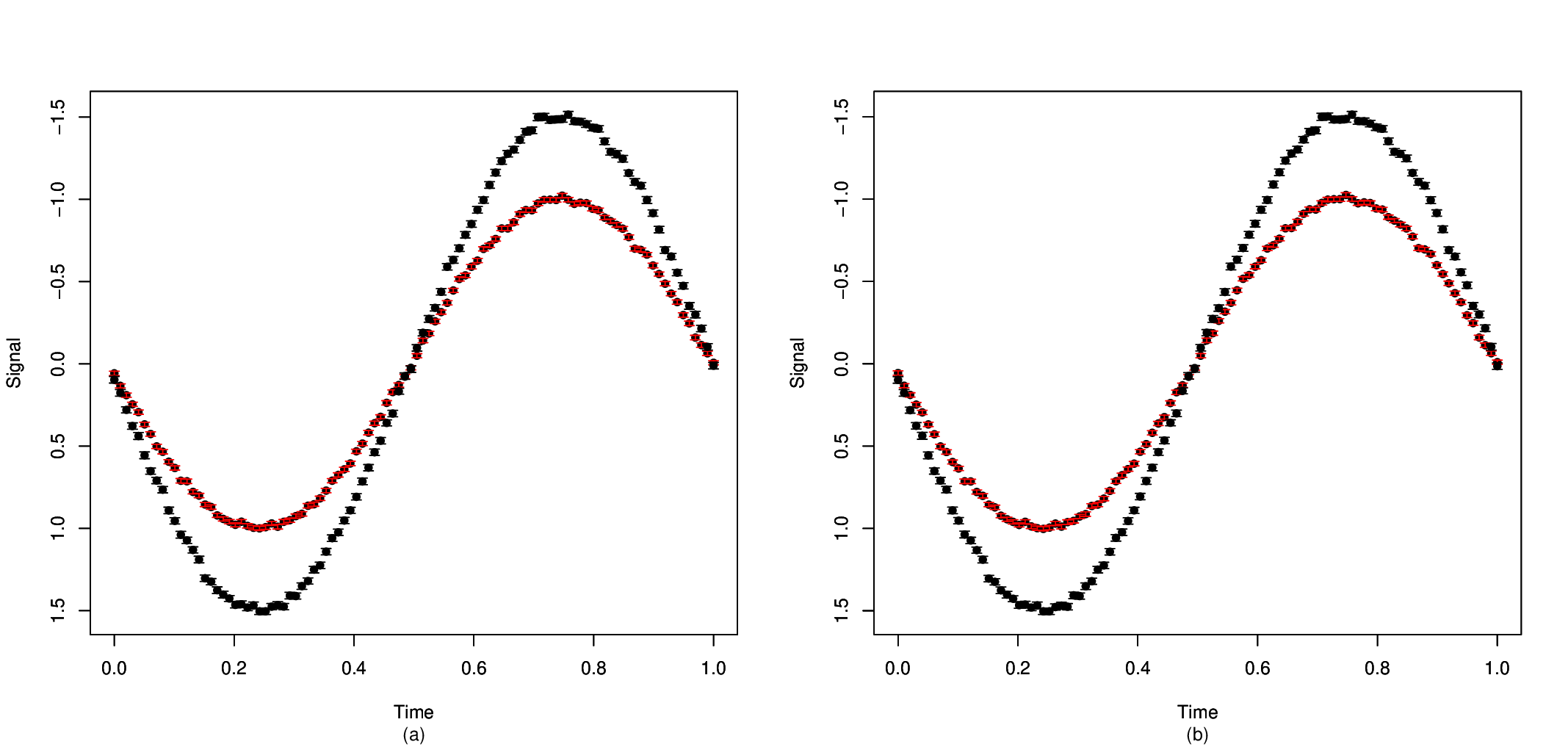}\\
		\caption{(a) Simulated average light curves (with standard error), generated using sine signal with added noise and outliers, of two groups each containing 1,000 LCs with respective amplitudes 1 (red) and 1.5 (black), (b) average light curves (with standard error) of two clusters, resulted in $k$-medoids clustering with CID, consisting of 1,004 (red) and 996 (black)	LCs.}\label{f1}
	\end{figure}

\begin{figure}
		\centering
		\includegraphics[width=1\textwidth]{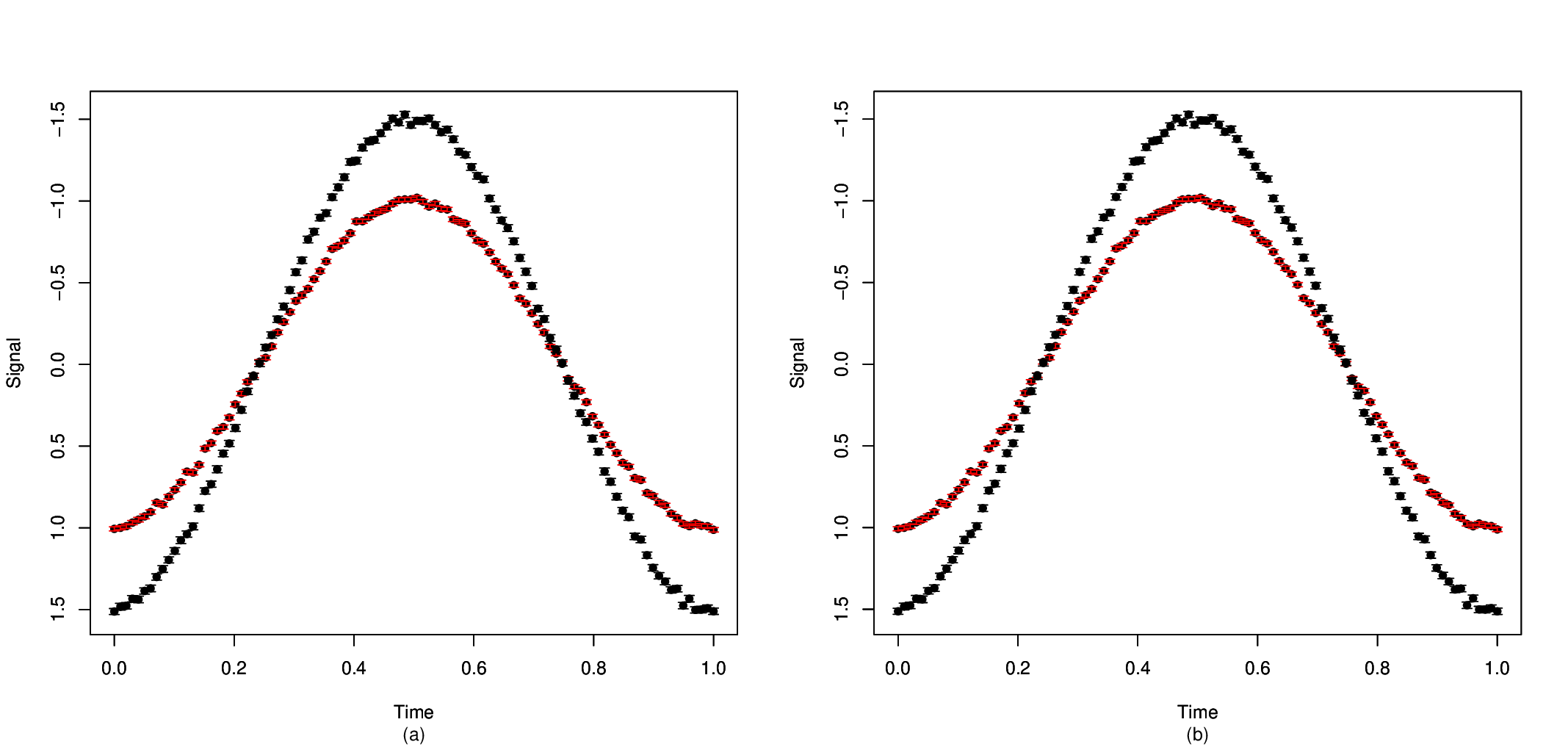}\\
		\caption{(a) Simulated average light curves (with standard error), generated using cosine signal with added noise and outliers, of two groups each containing 1,000 LCs with respective amplitudes 1 (red) and 1.5 (black), (b) average light curves (with standard error) of two clusters, resulted in $k$-medoids clustering with CID, consisting of 1,003 (red) and 997 (black) LCs.}\label{f2}
	\end{figure}

	We show the performance of $k$-medoids clustering with CID in exploring inherent groups of astronomical objects based on their LCs through simulation study. Here we generate LCs using a periodic signal contaminated with noise and outliers at 100 evenly spaced time points on [0,1] (see, for details, Thieler et al. 2013, 2016 and references therein). In the first case, we consider a complete cycle of the sine signal contaminated with signal-to-noise ratio $=3$, wherein 90\% of the noise is related to measurement accuracies and 10\% is white noise, as well as $10\%$ outliers added to the measurement accuracies. It simulates 1,000 LCs from each of two groups having respective amplitudes 1 and 1.5, whose group-wise average LCs are drawn in Fig.~\ref{f1}(a). Secondly, we generate LCs in the same way using a complete cycle of the cosine signal, whose group-wise average LCs are plotted in Fig.~\ref{f2}(a). Then, in each case we combine the generated LCs from two groups and perform our method on them which reveals the two existing clusters of the data in terms of ASW (see, Table~\ref{t1}). The average LCs of the two resulting clusters are shown in Fig.~\ref{f1}(b) for the first case and  in Fig.~\ref{f2}(b) for the latter. Our approach successfully distinguishes between the closely related inherent groups of noisy and outlier affected data in both the cases with $0.2\%$ and $0.15\%$ misclassification respectively. It is observed that the almost identical appearances between Figs.~\ref{f1}(a) and~\ref{f1}(b) and between Figs.~\ref{f2}(a) and~\ref{f2}(b) validate that the proposed clustering method is significantly efficient in identifying natural clustering under the considered situations with very low misclassification chances.
	
	\section{Result and discussion}\label{RAD}
\begin{figure}
		\centering
		\includegraphics[width=1\textwidth]{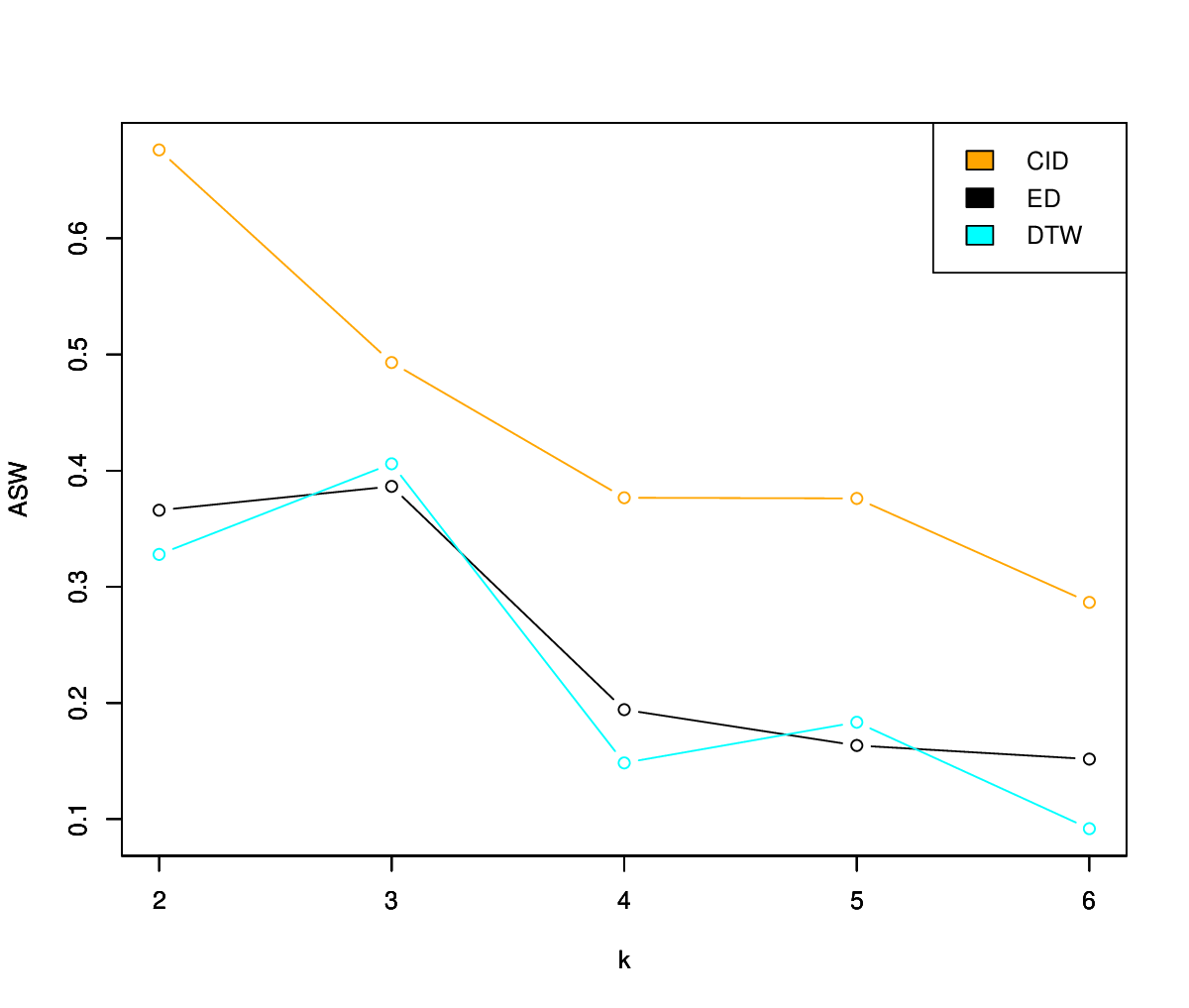}\\
		\caption{The average silhouette width (ASW) for different number of clusters ($k$) corresponding to three different distance measures in combination with $k$-medoids clustering method. The circles indicate values of ASW corresponding to a value of $k$.}\label{f3}
	\end{figure}

\begin{figure}
		\centering
		\includegraphics[width=1\textwidth]{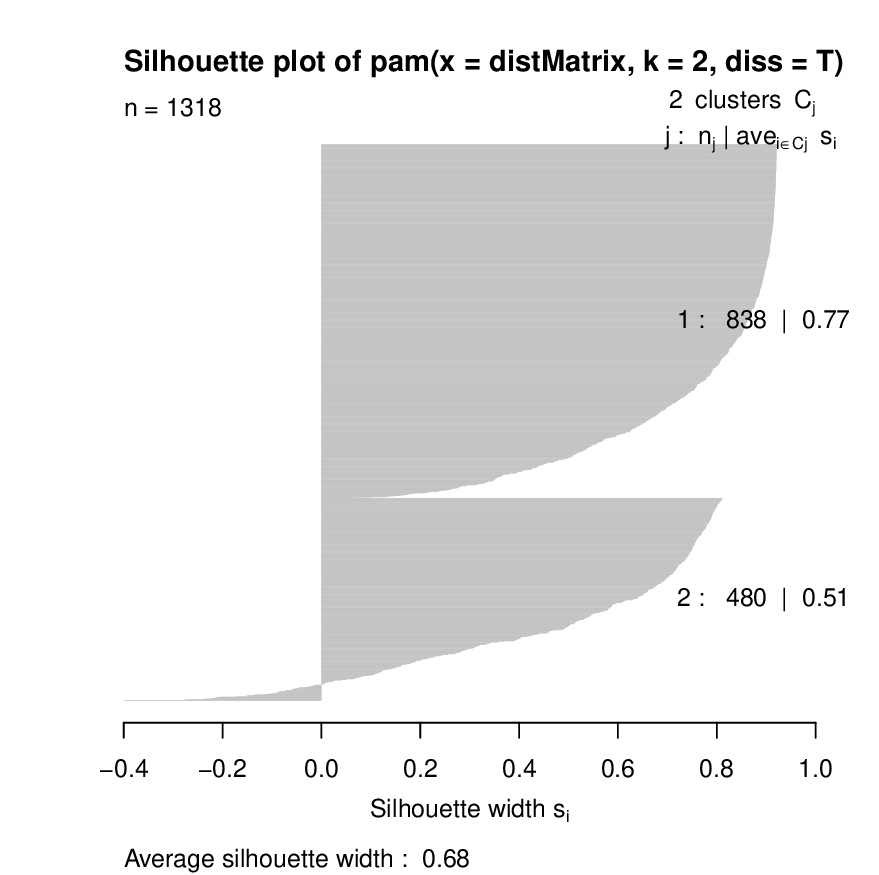}\\
		\caption{The silhouette plot gives a graphical representation of silhouette width of each of the light curves belonging to individual clusters, resulted from the $k$-medoids clustering with CID for $k=2$. The grey shade indicates the silhouette width of a light curve, arranged in descending order (from top to bottom) for individual clusters. The average silhouette width for two clusters k1, k2 and the whole data set of respective sizes of 838, 480 and 1318 are computed as 0.77, 0.51 and 0.68, respectively.}\label{f4}
	\end{figure}

\begin{figure}
		\centering
		\includegraphics[width=1\textwidth]{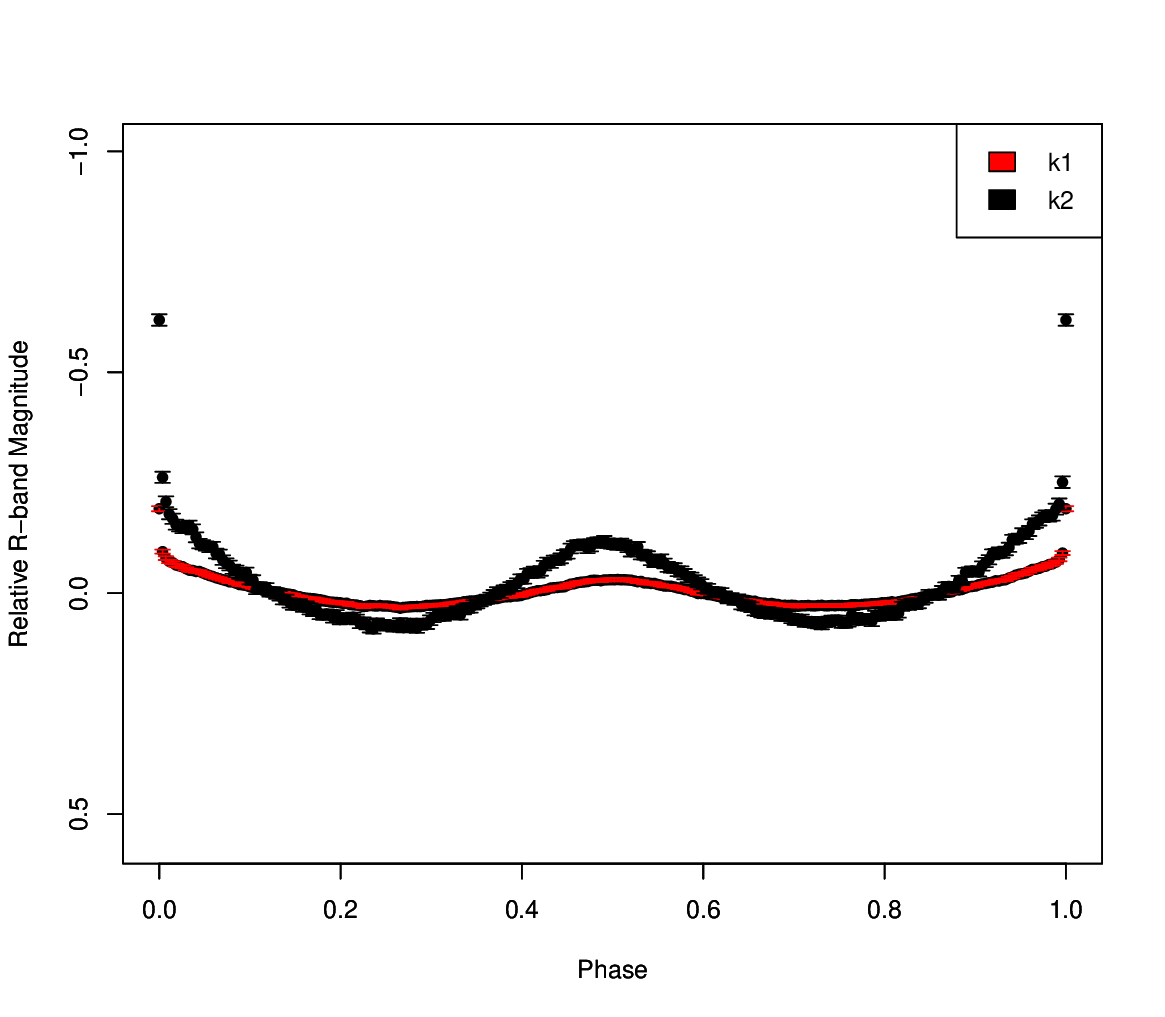}\\
		\caption{Template average light curves, with standard error, of two clusters k1 and k2 obtained from $k$-medoids clustering with CID.}\label{f5}
	\end{figure}

\begin{figure}
		\centering
		\includegraphics[width=1\textwidth]{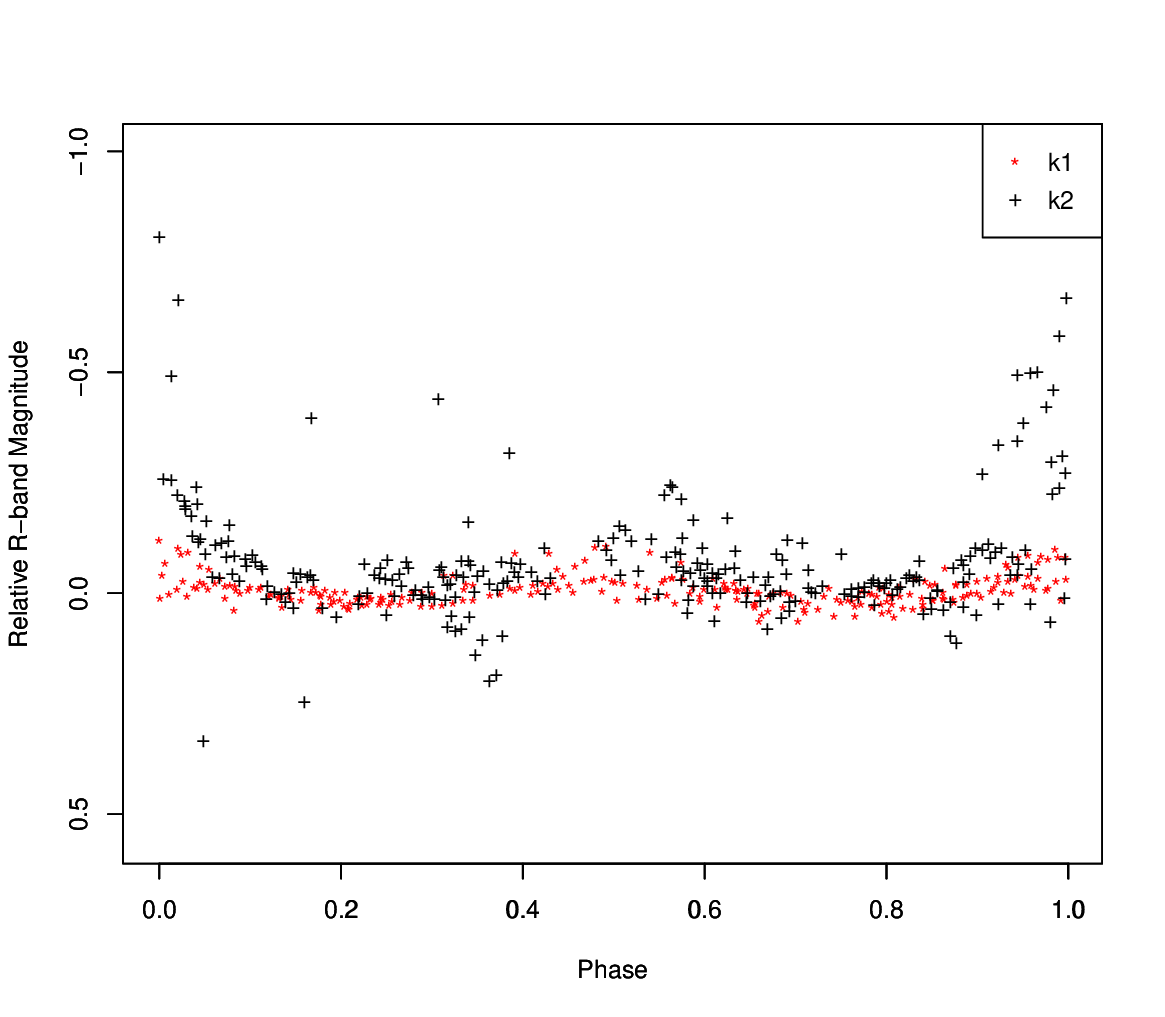}\\
		\caption{Template medoid light curves, corresponding to the stars with ID V-1221 and V-1138 (see, Miller et al. 2010), of two clusters k1 and k2 respectively obtained from $k$-medoids clustering with CID.}\label{f6}
	\end{figure}

\begin{figure}
		\centering
		\includegraphics[width=1\textwidth]{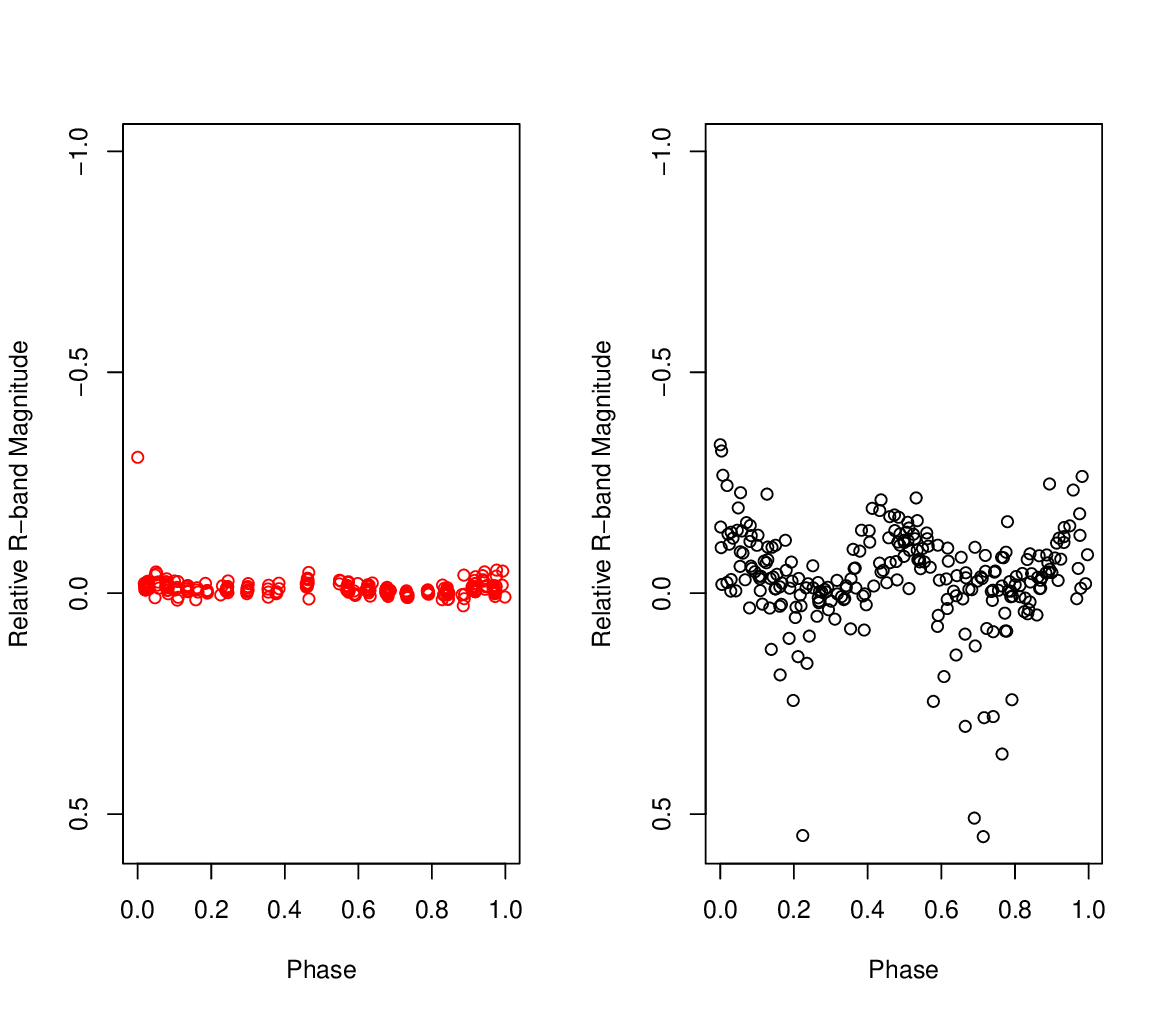}\\
		\caption{A pair of representative light curves, which are the observed light curves corresponding to the stars with ID V-94 and V-384 (see, Miller et al. 2010), of two clusters k1 (left) and k2 (right) respectively obtained from $k$-medoids clustering with CID.}\label{f7}
	\end{figure}

\begin{figure}
		\centering
		\includegraphics[width=1\textwidth]{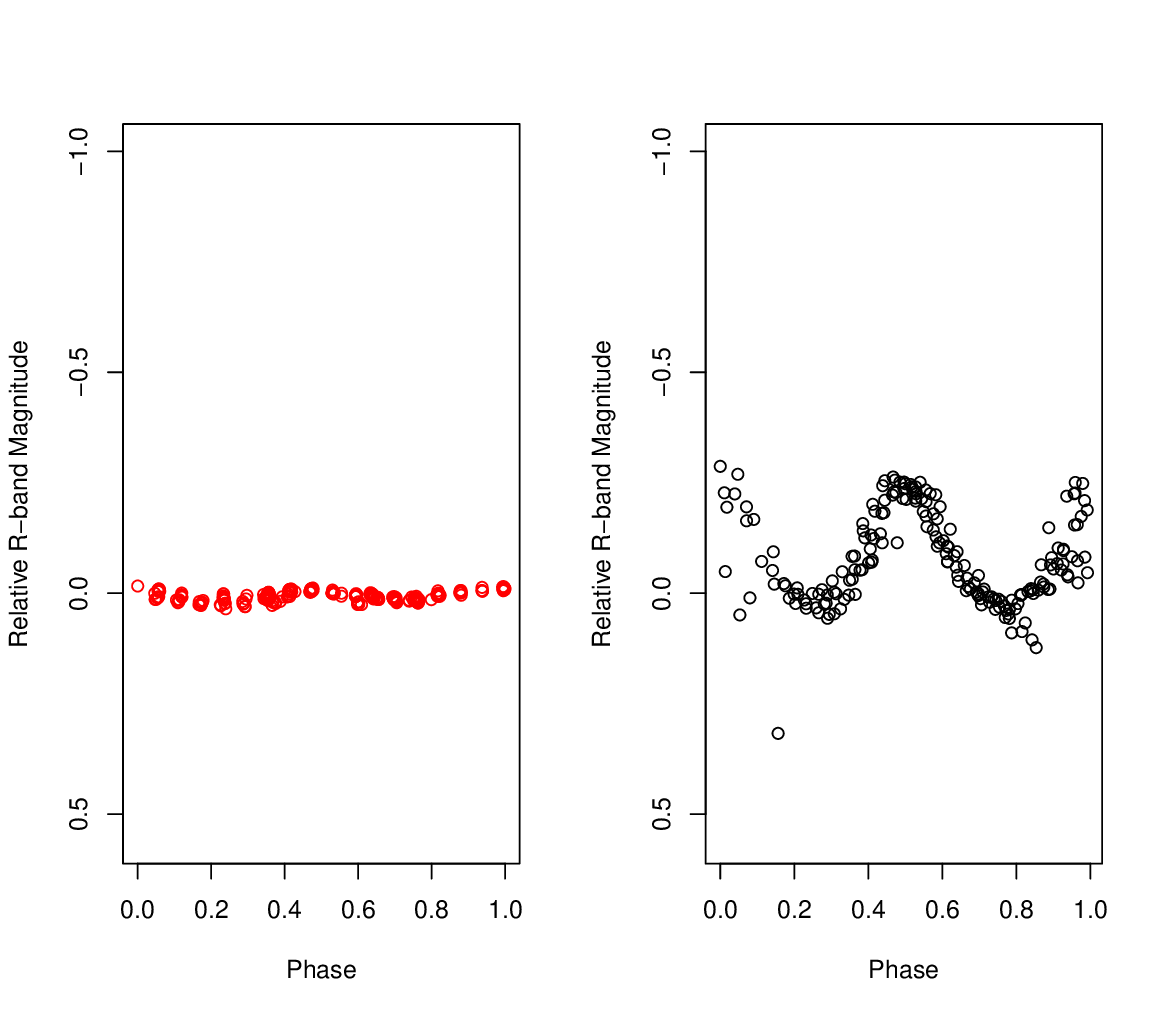}\\
		\caption{Another pair of representative light curves, which are the observed light curves corresponding to the stars with ID V-334 and V-817 (see, Miller et al. 2010), of two clusters k1 (left) and k2 (right) respectively obtained from $k$-medoids clustering with CID.}\label{f8}
	\end{figure}

	We compare the $k$-medoids clustering results obtained through CID, DTW and ED in terms of ASW and show that CID outperforms the other two (see, Fig. \ref{f3}) resulting in $k=2$ (see, Table \ref{t2}, which also verifies $k=2$ by the connectivity measure) with two clusters, denoted by k1 and k2, of respective sizes 838 and 480. The corresponding
	silhouette plot (Fig. \ref{f4}) displays the tightness of individual clusters and the separation between two clusters are significant. ASW for k1, k2 and for the whole data set of size 1318 are computed as 0.77, 0.51 and 0.68, respectively, show that the data is quite well clustered. This classification, irrespective of their subjective one (see, Table \ref{t3}), has template LCs in Figs. \ref{f5} and \ref{f6}. Cluster-wise representative LCs (i.e. two sets of observed LCs) in Figs. \ref{f7} and \ref{f8} indicate the similarity with their template LCs, whereas the average properties of two clusters are reported in Table \ref{t4}. 
	
	We compare our method with other existing unsupervised classifiers, where the DBSCAN method applied to the LCs with distance measure CID and parameters $\epsilon=0.5$, $MinPts=5$ fails to reveal the inherent clustering and results in only one group of size 253 with 1,065 LCs assigned as noise objects. Even if we consider the noise objects as a separate group (Kochoska et al. 2017), then also it gives a poor discrimination with ASW = 0.24. Again, we transform the LCs into two-dimensional data through $t$-SNE, where the distance between the LCs is measured by CID. Then DBSCAN $(\epsilon=0.5,MinPts=5)$ method, with distance measure ED performed on this transformed data, also fails to identify the inherent groups and gives five scattered clusters of sizes 5 to 7 and 1291 noise objects. However, our method which is robust against noise and outliers, successfully exposes the clusters from the LCs with significant accuracy in terms of ASW (see, Table \ref{t2} and Fig. \ref{f4}). We check superiority and robustness of our method by comparing it with $k$-means clustering, applied to linear features (Modak et al. 2018) extracted from the LCs in terms of the first ten principal components describing more than 80\% variation in the LCs, which also hints at two optimal groups of the variable stars (see, Table \ref{t5}).
	
\begin{figure}
		\centering
		\includegraphics[width=1\textwidth]{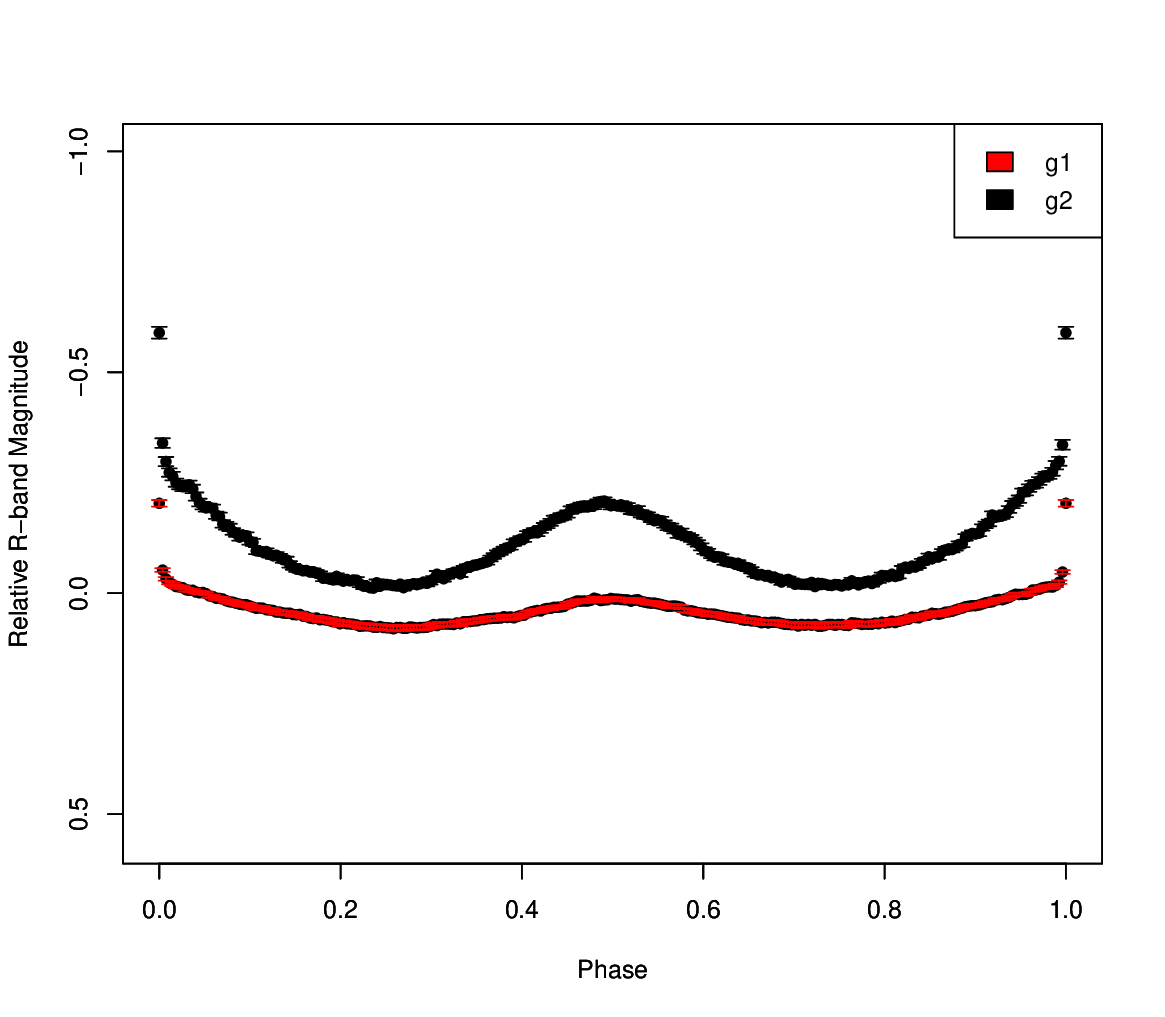}\\
		\caption{Template average light curves, with standard error, of two clusters g1 and g2 containing the distinct light curves obtained from fuzzy $k$-means clustering.}\label{f9}
	\end{figure}

	The morphologies of variable stars change continuously, so it may not always be possible to find distinctly separated clusters of the stars from their LCs. Therefore, we verify the distinction between the clusters by fuzzy $k$-means clustering, wherein the stars with $k=2$ whose LCs share potentially common properties of the groups can be classified into both the clusters. It gives two distinct groups of 811 and 427 LCs, whereas 80 LCs cannot be classified distinctly as their membership values are very close for the two groups. Hence, we eliminate these 80 LCs and obtain two groups, say g1 and g2, which contain only the non-overlapping LCs. Comparison between Tables \ref{t4} and \ref{t6} and Figs. \ref{f5} and \ref{f9} show the similarity between groups k1 and g1 and groups k2 and g2. It supports the distinction between the two groups obtained by our method. Also, the clustering with groups g1 and g2 has ASW = 0.28, which clearly shows that our method leads to significantly better clustering. Therefore, further astrophysical analyses of the groups are carried out based on the results obtained from $k$-medoids clustering with CID.
	
\begin{figure}
		\centering
		\includegraphics[width=1\textwidth]{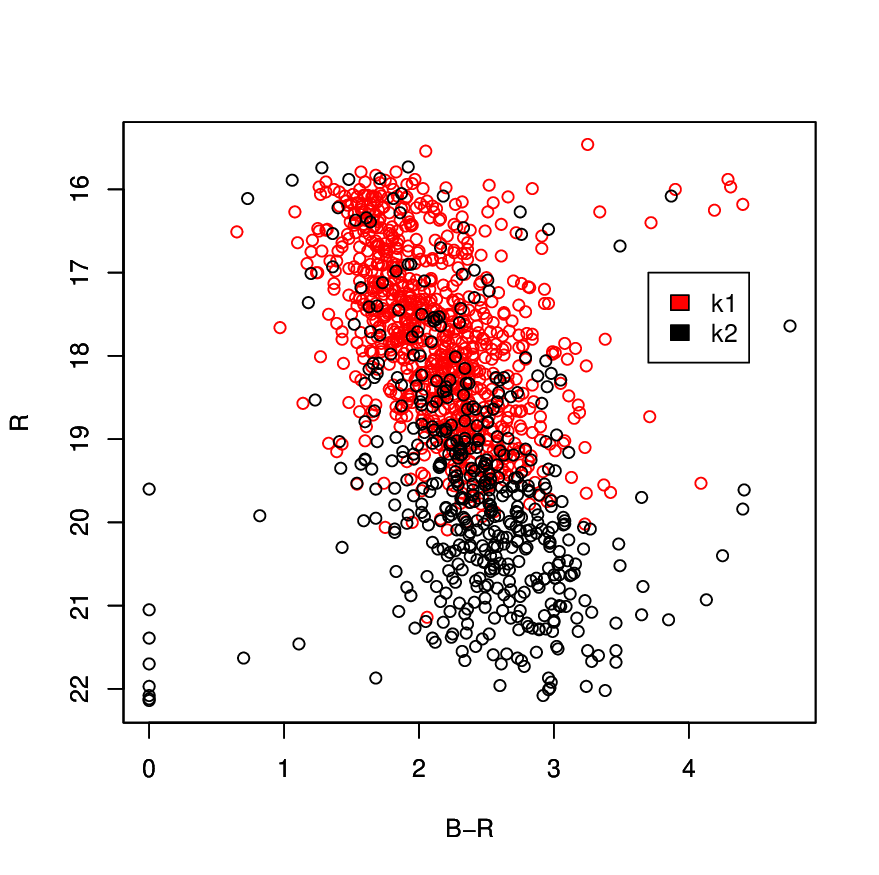}\\
		\caption{Color-magnitude diagram of the stars clustered in two groups k1 and k2 through $k$-medoids clustering method with CID.}\label{f10}
	\end{figure}

\begin{figure}
		\centering
		\includegraphics[width=1\textwidth]{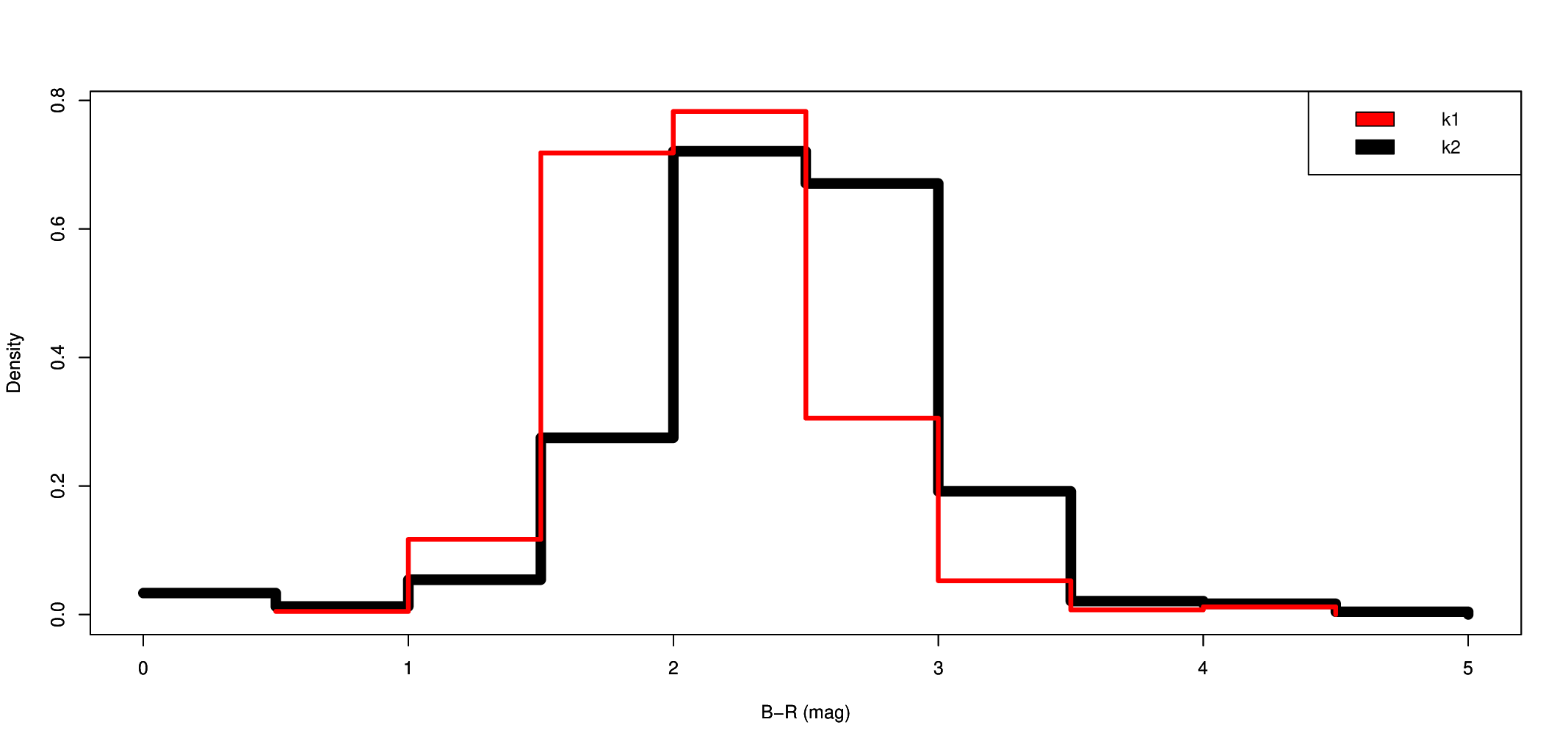}\\
		\caption{Histograms of B-R color index for the stars clustered in two groups k1 and k2 through $k$-medoids clustering method with CID.}\label{f11}
	\end{figure}

\begin{figure}
		\centering
		\includegraphics[width=1\textwidth]{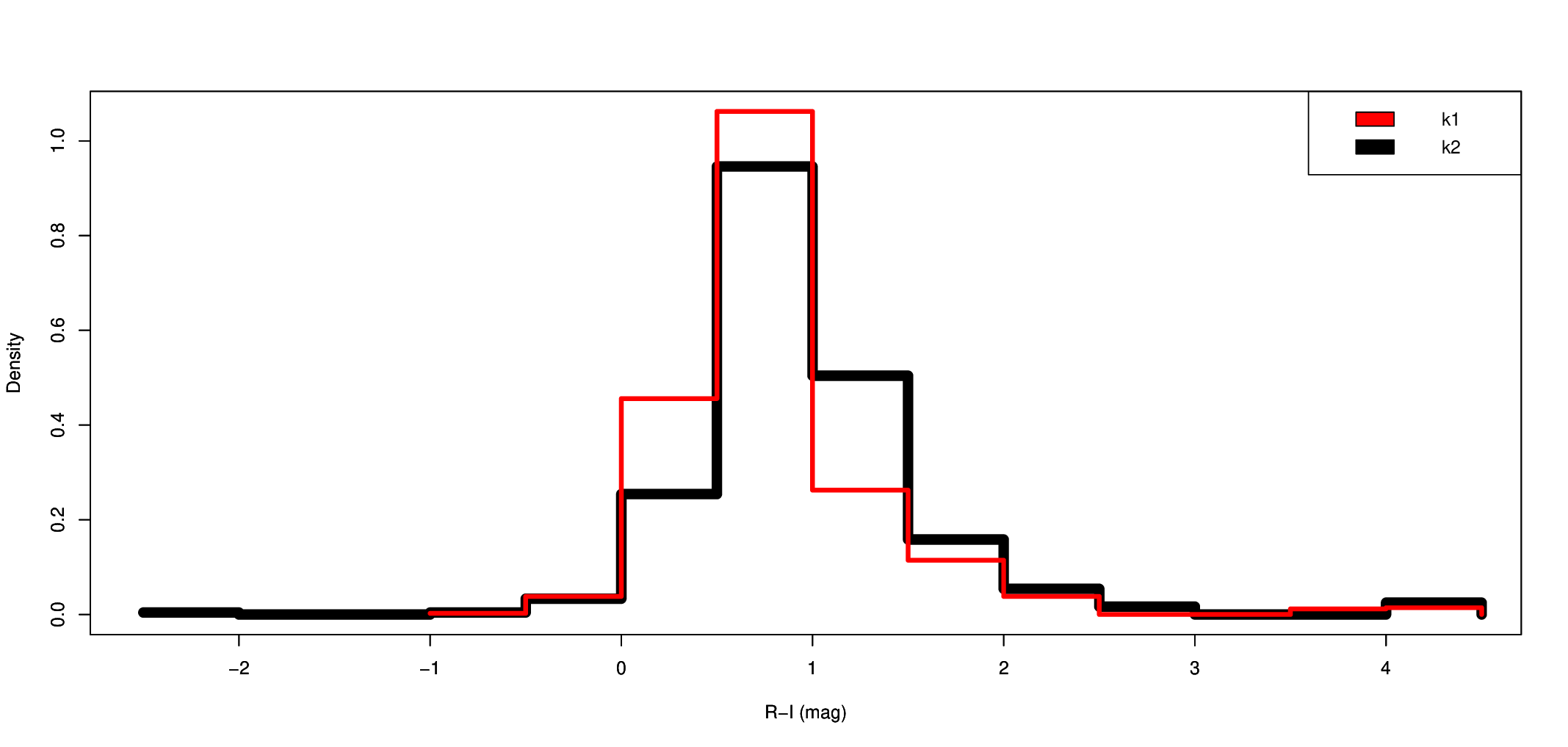}\\
		\caption{Histograms of R-I color index for the stars clustered in two groups k1 and k2 through $k$-medoids clustering method with CID.}\label{f12}
	\end{figure}

	Figs. $\ref{f5}-\ref{f8}$ and Table \ref{t4} indicate that the LCs in k1 have less variation between the two minima and larger average time period compared to those in k2. These suggest k1 systems consist of stars which form a more or less detached or semidetached system. Also the depths of the two minima of LCs for k1 are smaller compared to those for k2. This indicates k1 systems have a less massive secondary, whereas the masses are comparable for k2 systems. The color-magnitude diagram (Fig. \ref{f10}), the color histograms (Figs. \ref{f11} and \ref{f12}) and Table \ref{t4} show that k1 systems are bluer, i.e. have higher temperature than k2 systems and consist of stars with unequal mass, whereas the systems in k2 are redder and consist of stars with comparable mass. So k1 and k2 systems respectively belong to early and late spectral types.
	
	In this regard we discuss some of the recent works on classification of Es. Sarro et al. (2006) classified 81 Es using Bayesian model-based neural networks and resulted in groups which have a high degree of superposition with respect to properties like mass, period, separation. However, our work is more robust based on a nonparametric classification scheme which does not adopt any model assumptions, and there is a well-defined distinction between the resulting groups both in terms of the LCs (Figs. \ref{f5}--\ref{f8}) and the average properties, e.g. the period is almost double in k1 compared to k2 (see, Table \ref{t4}). Sarro et al. (2006) also concluded that classification based on the LCs, like we perform, is always better than classification with respect to the variables derived from the LCs. Malkov et al. (2007) classified 6330 Es on the basis of their observable variables, but their classification is subjective and restricted by several assumptions unlike our method. Pr\v{s}a et al. (2008) classified Es based on the variables, derived from 10,000 synthetic and 50 real LCs, by artificial neural network (ANN) method with presumed five classes, whereas our method is applied to the observed LCs with the number of classes found scientifically. Matijevi\v{c} et al. (2012) used dimension reduction technique based on LLE algorithm and found that the projection onto a two-dimensional space can preserve the local geometry. This is somewhat consistent with our findings of two groups of Es, but finally their groups reduced to a single variable equivalent to `detachedness' of the binaries. Kirk et al. (2016) classified about 2,00,000 Es based on their observed LCs by LIE method, but they also assumed the number of classes as prerequisite. In Kochoska et al. (2017); Mowlavi et al. (2017) and S\"{u}veges et al. (2017), the classification error is
	around 10\% mainly due to the similarity of LCs originating from
	different physical systems. In particular, Kochoska et al. 2017 have found
	four groups of which the first two and the last two have similar Kepler polyfit
	primary depths indicating merely two significant groups of Es which supports our results of two groups.
	\section{Conclusion}
	We have classified 1,318 variable stars in the Galaxy which lie primarily along the spiral arms. To overcome the deficiencies of the subjective classification method  (Miller et al. 2010), $k$-medoids clustering with CID is applied to the LCs, which gives rise to two physically interpretable groups of Es and indicates there is no separate group of pulsating stars in the present data set. The accuracy of the resulting clustering is significant in terms of the ASW and having outperformed the established methods in astronomical literature, our approach is a strong competitor for future classification of new variable stars based on their LCs.
	
	\section{Acknowledgements}
	The authors would like to thank the Editor-in-chief and an anonymous associate editor for encouraging the present work on Astrostatistics and two anonymous referees for their intriguing inquiries which helped the authors to present the results in a more convincing way.
	\clearpage
	\begin{table}
		\caption{The average silhouette width (ASW$\times10^2$) for different number of clusters ($k$), from $k$-medoids clustering with CID, for two simulated data sets with ASW$_{sin}$ and ASW$_{cos}$ respectively corresponding to the sine and cosine signal generated data sets}
		\begin{center}
			\begin{tabular}{c c c}
				\hline\hline
				\noalign{\vskip .05in}
				$k$ & ASW$_{sin}$ & ASW$_{cos}$\\ [0.5 ex]
				\hline
				\noalign{\vskip .05in}
				2 & 37.758 & 37.611\\
				3 & 25.372 &25.162\\
				4 &  0.901 &1.095\\
				5 &  0.872 &1.062\\
				6 &  0.705 &0.877\\
				[0.5 ex]
				\hline
			\end{tabular}
		\end{center}
		\label{t1}
	\end{table}
	\clearpage
	\begin{table}
		\caption{The average silhouette width (ASW$\times10^2$) and the connectivity for different number of clusters ($k$) from $k$-medoids clustering with CID}
		\begin{center}
			\begin{tabular}{c c c}
				\hline\hline
				\noalign{\vskip .05in}
				$k$ & ASW & Connectivity\\ [0.5 ex]
				\hline
				\noalign{\vskip .05in}
				2 & 67.610 &149.690\\
				3 & 49.309 &246.292\\
				4 & 37.666 &395.906\\
				5 & 37.608 &421.292\\
				6 & 28.659 &513.438\\
				[0.5 ex]
				\hline
			\end{tabular}
		\end{center}
		\label{t2}
	\end{table}
	\clearpage
	\begin{table}
		\caption{Membership of subjective types in two groups k1 and k2}
		\begin{center}
			\begin{tabular}{c c  c}
				\hline\hline
				\noalign{\vskip .05 in}
				Type &k1 & k2\\[.5 ex]
				\hline
				\noalign{\vskip .05 in}
				EA&43&44\\
				EB&  84& 24\\
				EW&234&260\\
				PUL&99&19\\
				EA:&15&31\\
				EB:&165&50\\
				EW:&7&14\\
				PUL:&145&23\\
				CV:&1&0\\
				EA/EB& 6&4\\
				EW/EA&2& 4\\
				EW/EB&8&4\\
				EB/PUL&11&1\\
				DCEP/PUL&9&0\\
				CV/PUL&9&2\\[.5 ex]
				\hline
				\noalign{\vskip .05 in}
				Total&838&480\\[.5 ex]
				\hline
				\noalign{\vskip .05 in}
				Note: An uncertain type is followed by a
				colon and\\ an ambiguous type is given with a slash.
			\end{tabular}
		\end{center}
		\label{t3}
	\end{table}
	\clearpage
	\begin{table}
		\caption{Average values (with standard error) of the variables for two clusters k1 and k2 obtained from $k$-medoids clustering with CID}
		\begin{center}
			\tiny
			\begin{tabular}{c c c c c c c c}
				\hline\hline
				\noalign{\vskip .05 in}
				Name of  & No. of  &P & R                & B                & I                & B-I               & R-I \\ [0.5ex]
				cluster  & members &(day) &(mag) &(mag) &(mag) &(mag) &(mag)\\[0.5ex]
				\hline
				\noalign{\vskip .05in}
				k1     &838    &2.816$\pm$0.045       & 17.783$\pm$0.036 & 19.917$\pm$0.045 & 16.968$\pm$0.037 & 2.948$\pm$0.026 & 0.814$\pm$0.020 \\
				k2     & 480   &1.400$\pm$0.125       & 19.636$\pm$0.062 & 22.046$\pm$0.075 & 18.657$\pm$0.061 & 3.389$\pm$0.040 & 0.978$\pm$0.028
				\\[.5ex]
				\hline
			\end{tabular}
		\end{center}
		\label{t4}
	\end{table}
	\clearpage
	\begin{table}
		\caption{The average silhouette width (ASW$\times10^2$) for different number of clusters ($k$) from $k$-means clustering}
		\begin{center}
			\begin{tabular}{c c}
				\hline\hline
				\noalign{\vskip .05in}
				$k$  & ASW\\ [0.5 ex]
				\hline
				\noalign{\vskip .05in}
				2  & 52.426\\
				3  &  47.398\\
				4  & 35.575\\
				5  &  31.107 \\
				6  & 23.574 \\
				[0.5 ex]
				\hline
			\end{tabular}
		\end{center}
		\label{t5}
	\end{table}
	\clearpage
	\begin{table}
		\caption{Average values (with standard error) of the variables for two clusters g1 and g2 containing the distinct light curves, obtained from fuzzy $k$-means clustering}
		\begin{center}
			\tiny
			\begin{tabular}{c c c c c c c c}
				\hline\hline
				\noalign{\vskip .05 in}
				Name of  & No. of  &P & R                & B                & I                & B-I               & R-I \\ [0.5ex]
				cluster  & members &(day) &(mag) &(mag) &(mag) &(mag) &(mag)\\[0.5ex]
				\hline
				\noalign{\vskip .05in}
				g1     &811  & 2.896$\pm$0.135       & 17.955$\pm$0.044 & 20.077$\pm$0.053 & 17.095$\pm$0.044 & 2.981$\pm$0.028 &  0.859$\pm$0.021 \\
				g2     &427   &1.375$\pm$0.108       & 19.299$\pm$0.069 & 21.724$\pm$0.081 & 18.394$\pm$0.066 &  3.330$\pm$0.043 &  0.905$\pm$0.031
				\\[.5ex]
				\hline
			\end{tabular}
		\end{center}
		\label{t6}
	\end{table}
	\clearpage
	\clearpage
	\clearpage
	\clearpage
	\clearpage
	\clearpage
	\clearpage
	\clearpage
	\clearpage
	\clearpage
	\clearpage
	\clearpage
	
\end{document}